\documentclass{aa}

\usepackage[varg]{txfonts}

\usepackage{natbib}
\bibpunct{(}{)}{;}{a}{}{,} 

\title{Main sequence of star forming galaxies beyond the \textit{Herschel}\thanks{{\it Herschel} is an ESA space observatory with science instruments provided by European-led Principal Investigator consortia and with important participation from NASA.} confusion limit}
\titlerunning{The Main Sequence beyond the \textit{Herschel} Confusion Limit}
\author{W.~J.~Pearson\inst{\ref{inst:SRON}, \ref{inst:Kapteyn}},
L.~Wang\inst{\ref{inst:SRON}, \ref{inst:Kapteyn}},
P.~D.~Hurley\inst{\ref{inst:Sussex}},
K.~Ma\l{}ek\inst{\ref{inst:LAM} ,\ref{inst:NCBJ}},
V.~Buat\inst{\ref{inst:LAM}},
D.~Burgarella\inst{\ref{inst:LAM}},
D.~Farrah\inst{\ref{inst:Virginia}},
S.~J.~Oliver\inst{\ref{inst:Sussex}},
D.~J.~B.~Smith\inst{\ref{inst:Herts}},
F.~F.~S.~van~der~Tak\inst{\ref{inst:SRON}, \ref{inst:Kapteyn}}
}
\institute{SRON Netherlands Institute for Space Research, Landleven 12, 9747 AD, Groningen, The Netherlands\label{inst:SRON}\\\email{w.j.pearson@sron.nl}
\and Kapteyn Astronomical Institute, University of Groningen, Postbus 800, 9700 AV Groningen, The Netherlands\label{inst:Kapteyn}
\and Astronomy Centre, Department of Physics and Astronomy, University of Sussex, Falmer, Brighton BN1 9QH, UK\label{inst:Sussex}
\and Aix-Marseille Universit\'{e}, CNRS, LAM (Laboratorie d'Astrophysique de Marseille) UMR 7326, 13388 Marseille, France\label{inst:LAM}
\and National Centre for Nuclear Research, ul. Hoza 69, 00-681 Warszawa, Poland\label{inst:NCBJ}
\and Department of Physics, Virginia Tech, Blacksburg, VA 24061, USA\label{inst:Virginia}
\and Centre for Astrophysics, Science \& Technology Research Institute, University of Hertfordshire, Hatfield, Herts, AL10 9AB, UK\label{inst:Herts}}
\authorrunning{W.~J.~Pearson et al.}

\date{Received 13 February 2018 /
Accepted 27 March 2018}
\abstract{Deep far-infrared (FIR) cosmological surveys are known to be affected by source confusion, causing issues when examining the main sequence (MS) of star forming galaxies. In the past this has typically been partially tackled by the use of stacking. However, stacking only provides the average properties of the objects in the stack.}
{This work aims to trace the MS over $0.2\leq z <6.0$ using the latest de-blended \textit{Herschel} photometry, which reaches $\approx10$ times deeper than the 5$\sigma$ confusion limit in SPIRE. This provides more reliable star formation rates (SFRs), especially for the fainter galaxies, and hence a more reliable MS.}
{We built a pipeline that uses the spectral energy distribution (SED) modelling and fitting tool CIGALE to generate flux density priors in the \textit{Herschel} SPIRE bands. These priors were then fed into the de-blending tool XID+ to extract flux densities from the SPIRE maps. In the final step, multi-wavelength data were combined with the extracted SPIRE flux densities to constrain SEDs and provide stellar mass (M$_{\star}$) and SFRs. These M$_{\star}$ and SFRs were then used to populate the SFR-M$_{\star}$ plane over $0.2 \leq z < 6.0$.}
{No significant evidence of a high-mass turn-over was found;  the best fit is thus a simple two-parameter power law of the form $\log(\mathrm{SFR})=\alpha[\log(\mathrm{M}_{\star})-10.5]+\beta$. The normalisation of the power law increases with redshift, rapidly at $z\lesssim1.8$, from $0.58\pm0.09$ at $z\approx0.37$ to $1.31\pm0.08$ at $z\approx1.8$. The slope is also found to increase with redshift, perhaps with an excess around $1.8\leq z<2.9$.}
{The increasing slope indicates that galaxies become more self-similar as redshift increases. This implies that the specific SFR of high-mass galaxies  increases with redshift, from 0.2 to 6.0, becoming closer to that of low-mass galaxies. The excess in the slope at $1.8\leq z<2.9$, if present, coincides with the peak of the cosmic star formation history.}


\begin{document}

\maketitle

\section{Introduction} \label{sec:intro}
It has been observed that there is a strong correlation between the stellar mass (M$_\star$) and the star formation rate (SFR) for the majority of star forming galaxies (SFG) \citep[e.g.][]{2004MNRAS.351.1151B, 2007ApJ...660L..43N, 2007AA...468...33E, 2014ApJS..214...15S, 2016ApJ...817..118T}. This has become known as the main sequence (MS) of star forming galaxies.

The MS is notable due to its tight correlation; the scatter of the SFR-M$_{\star}$ relation has been found to be approximately 0.3~dex. This low scatter has been observed to exist for approximately the last 10~Gyr and has been found to be independent of M$_{\star}$ \citep{2012ApJ...754L..29W, 2014ApJS..214...15S, 2016ApJ...817..118T}. The consistency of the scatter is understood to be a result of every galaxy having star formation (SF) regulated by similar quasi-static processes \citep{2015ApJ...801...80L}, while the scatter arises as a result of minor fluctuations of the flow of material into galaxies \citep{2016MNRAS.457.2790T, 2017MNRAS.464.2766M}. Galaxies move to the upper MS as the gas is compacted in the centre of the galaxy which triggers SF. As the central gas is depleted, but before the galaxy moves above the MS, the SFR reduces and the galaxy falls to the lower MS. New gas then falls into the galaxy, replenishing the central gas reservoir before the galaxy becomes quiescent. This cycle repeats until the galaxy's gas replenishment time is longer than the depletion time and the galaxy becomes quiescent \citep{2016MNRAS.457.2790T}.

Two different schools of thought exist over the shape the MS takes. Some studies find the relation between SFR and stellar mass is a simple power law \citep[e.g.][]{2014ApJS..214...15S}, i.e. the log of the SFR increases with the log of the stellar mass as
\begin{equation}\label{eqn:linear}
\log(\mathrm{SFR}/\mathrm{M}_{\odot} \mathrm{yr}^{-1}) = \alpha \log(\mathrm{M}_{\star}/\mathrm{M}_{\odot}) + \beta,
\end{equation}
where $\alpha$ is the slope and $\beta$ is the normalisation. However, other studies find there is a high-mass turn-over at log(M$_{\star}$/M$_{\odot}$) $\approx$ 10.5 with the slope of the MS being shallower above the turn-over than below \citep[e.g.][]{2015ApJ...801...80L, 2016ApJ...817..118T}. Recent studies have shown the two different forms of the MS may be a result of how the SFG population is separated from the quiescent galaxies (QG). \citet{2015MNRAS.453.2540J} have shown that using a SFG-QG cut that is stricter in selecting the SFG population, in this case a 4000~\text{\AA} break index value that is lower, will result in a straight MS, while a cut that is less strict forms a MS with a turn-over. The less strict cut leaves in high-mass, low SFR objects, which lower the mean SFR at high mass.

Regardless of the form of the MS, the normalisation is found to increase with redshift \citep[e.g.][]{2014ApJS..214...15S, 2015MNRAS.453.2540J, 2015A&A...575A..74S, 2016ApJ...817..118T}. This increasing normalisation is not surprising. As redshift increases, the fraction of cold gas available for star formation increases \citep{2010Natur.463..781T, 2011MNRAS.417.1510D, 2015ApJ...800...20G, 2016ApJ...820...83S}. Thus, with more gas available, more stars can form, raising the average SFR in all galaxies.  This could be counterbalanced by a lower SFR per gas mass (SFR/M$_{\mathrm{gas}}$) at higher redshift but existing works have shown that SFR/M$_{\mathrm{gas}}$ is either relatively constant or rises with redshift \citep{2010Natur.463..781T, 2016ApJ...820...83S}.

It has been argued that the slope of the MS should be unity for all mass ranges once the SFR has become stable in a galaxy \citep{2017ApJ...834...39P}. However, the slope of the MS is typically found to be lower than unity, with values between 0.4 and 1.0 \citep[e.g.][]{2012ApJ...754L..29W, 2014ApJS..214...15S, 2016ApJ...817..118T}. This discrepancy is believed to be the result of a combination of quenching and a reduction in the relative size of the cold gas reservoir as the M$_{\star}$ of a galaxy increases \citep{2017ApJ...834...39P}.

Far-infrared (FIR) emission is a key component for accurately determining the SFR of an object. Part of the ultraviolet (UV) emission from young stars heats dust in a galaxy, which then re-radiates in the infrared (IR). Observing in the UV would provide a direct measure of SFR, but could underestimate the total SFR as this absorption by dust obscures some of the UV emission \citep[e.g.][]{1999ApJ...521...64M, 2009ApJ...703..517D, 2017MNRAS.467.1360B, 2017MNRAS.466..861D}. As a result, observing IR emission in combination with UV emission is key to providing a complete picture of the SFR.

However, FIR surveys, such as those conducted with the ESA \textit{Herschel Space Observatory} \citep{2010AA...518L...1P} Spectral and Photometric Imaging Receiver \citep[SPIRE;][]{2010AA...518L...3G}, have  relatively poor resolution with respect to optical, and also have source confusion \citep[e.g.][]{2010AA...518L...5N,2012MNRAS.424.1614O,2017MNRAS.464..885H}. Previously, stacking was used to recover the flux densities of fainter sources \citep[e.g.][]{2015ApJ...807..141P, 2015A&A...575A..74S, 2016MNRAS.457.4179H, 2016A&A...587A.122A}. However, by its very nature, stacking only provides the average properties (e.g. mean, median)  of the objects in the stack\footnote{It is possible to recover the scatter of the objects in a stack. However, this requires strong assumptions on the distribution of the galaxies in the stack \citep{2015A&A...575A..74S, 2016A&A...592L...5W}.}. As a result, the properties of individual objects cannot be determined, which results in the loss of information about the wider properties of a galaxy population.

To  overcome confusion, it is necessary to de-blend the maps to generate individual flux density measurements for both faint and bright sources. For SPIRE, this was done with the De-blended  SPIRE Photometry algorithm \citep[DESPHOT;][]{2010MNRAS.409...48R, 2012MNRAS.419.2758R, 2014MNRAS.444.2870W}, amongst other techniques \citep[e.g.][]{2007PASP..119.1325L, 2010A&A...516A..43B, 2010AJ....139.1592K, 2013ApJ...779...32V, 2015A&A...582A..15M, 2015ApJ...798...91S, 2016MNRAS.460..765W}. DESPHOT misassigns flux densities when more than one source is within a beam, and is also  unable to realistically derive flux density errors of a given source \citep[see discussion in][]{2017MNRAS.464..885H}. To improve source de-blending, XID+ \citep[][see also Sect. \ref{subsec:xidp}]{2017MNRAS.464..885H} has been developed and subsequently expanded to improve flux density estimation by including more precise flux density priors \citep{2017A&A...603A.102P}.

With XID+, it has been shown that more than 95\% of blindly detected SPIRE 250~$\mu$m sources  with a flux density greater than 30~mJy  contain more than one object which contribute more than 10\% of the source's total flux density. At least 70\% of the flux density from these sources is assigned to the two brightest objects \citep{2016MNRAS.460.1119S}. This suggests that many current SPIRE catalogues have too much flux density assigned to their objects, and  any derived physical parameter that relies on the SPIRE emission, such as the total infrared luminosity or star formation rate (SFR), will be overestimated.

In this work, XID+ is used to de-blend the maps in the SPIRE bands, resulting in a catalogue of over 200\,000 objects with SPIRE flux density measurements in the COSMOS field, compared to approximately 32\,000 sources blindly detected at 250~$\mu$m \citep{2010MNRAS.409...48R, 2014MNRAS.444.2870W, 2015MNRAS.447.3548S}. These data, along with UV to \textit{Herschel} Photodetector Array Camera and Spectrometer \citep[PACS;][]{2010A&A...518L...2P} data from a multi-wavelength catalogue, are then used to examine the MS. 

The paper is structured as follows. Section \ref{sec:data} describes where the data  were collected. Section \ref{sec:method} explains the methodology and tools used to de-blend the SPIRE bands and find the MS. Section \ref{sec:results} explores the results of the analysis. Section \ref{sec:discussion} provides discussion, while Section \ref{sec:conc} provides a summary of the conclusions. Where necessary, Wilkinson Microwave Anisotropy Probe year 7 (WMAP7) cosmology \citep{2011ApJS..192...18K, 2011ApJS..192...16L} is followed, with $\Omega_{M}$ = 0.272, $\Omega_{\Lambda}$ = 0.728, and H$_{0}$ = 70.4~km~s$^{-1}$~Mpc$^{-1}$, to be consistent with CIGALE\footnote{\url{http://cigale.lam.fr/}} \citep[][see also Sect. \ref{subsec:cigale}]{2009AA...507.1793N}.

\section{Data sets} \label{sec:data}
For this work, the COSMOS field \citep{2007ApJS..172....1S} was chosen due to the wealth of multi-wavelength data available within its two square degree coverage. It also benefits from  \textit{Herschel} SPIRE data, which  is considered to be homogenous and of high quality \citep[e.g.][]{2017A&A...603A.102P}.

\begin{figure}
        \centering
        \includegraphics[width=0.45\textwidth]{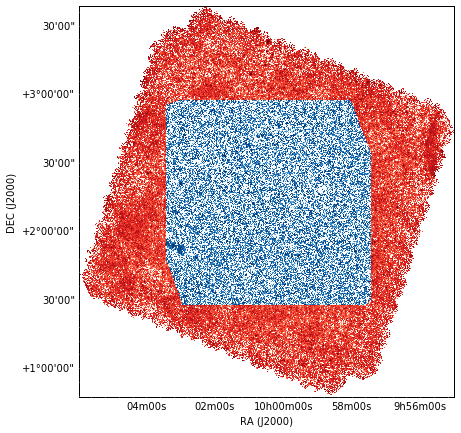}
        \caption{Image of the SPIRE 250~$\mu$m COSMOS coverage \citep[][7.84 deg$^{2}$, red]{2012MNRAS.424.1614O} with the overlayed MIPS 24~$\mu$m COSMOS coverage \citep[][2.23 deg$^{2}$, blue]{2007ApJS..172...86S}. The data used in this work were cut to match the MIPS 24~$\mu$m coverage.}
        \label{fig:cut}
\end{figure}

For the multi-wavelength data, the public COSMOS2015 catalogue \citep{2016ApJS..224...24L} was used with CIGALE to generate flux density priors in the 250~$\mu$m, 350~$\mu$m, and 500~$\mu$m SPIRE bands for XID+ (see Sect. \ref{subsec:cigale} and \ref{subsec:scipip}). COSMOS2015 contains photometric data in over 30 bands, narrow, medium, and broad, for approximately $1.2 \times 10^{6}$ objects in the COSMOS field. Here, we only use the bands that cover the UV to PACS 160~$\mu$m in CIGALE; a full list of the bands used can be found in Table \ref{table:bands}. The aperture photometry was converted to total photometry and corrected for foreground extinction following \citet{2016ApJS..224...24L}. As we wanted to use the Multi-band Imaging Photometer for \textit{Spitzer} 24~$\mu$m \citep[MIPS24;][]{2004ApJS..154...25R} data for the $\approx$40\,000 objects with MIPS24 data to help constrain the FIR SED, we used objects that only fall within the MIPS24 image in COSMOS. The MISP24 coverage falls within the SPIRE coverage; the requirements are as follows (see also Fig. \ref{fig:cut}):

\begin{equation} \label{eqn:cut}
        \begin{gathered}
        149.38^{\circ} \leq RA \leq 150.86^{\circ}\\
        1.46^{\circ} \leq Dec \leq 2.95^{\circ}\\
        Dec + (0.38 \times RA) < 59.51^{\circ}\\
        Dec - (2.49 \times RA) < -368.73^{\circ}\\
        Dec + (0.35 \times RA) > 53.91^{\circ}\\
        Dec - (2.66 \times RA) > -400.06^{\circ}
        \end{gathered}
\end{equation}

The latest SPIRE images from the \textit{Herschel} Database in Marseille\footnote{\url{http://hedam.lam.fr}}, the Data Release 4 maps from the \textit{Herschel} Multi-tiered Extragalactic Survey \citep{2012MNRAS.424.1614O}, were used in XID+ to extract the SPIRE flux densities. The 250~$\mu$m, 350~$\mu$m, and 500~$\mu$m band images have beam full widths at half maximum of 18.1$^{\prime\prime}$, 25.2$^{\prime\prime}$, and 36.6$^{\prime\prime}$ \citep{2010AA...518L...3G} and 5$\sigma$ confusion limits of 24.0, 27.5, and 30.5 mJy, respectively  \citep{2010AA...518L...5N}. 

\begin{table}
        \caption{Telescopes and associated bands that were used for CIGALE spectral energy distribution fitting from the COSMOS2015 catalogue.}
        \begin{center}
                \begin{tabular}{ll}
                \hline
                Telescope & Band(s)\\
                \hline
                GALEX & FUV, NUV\\
                CFHT & u\\
                Subaru & B, V, r, i+, z++, IB427, \\
                 & IB464, IA484, IB505, IA527, IB574,\\
                 & IA679, IB709, IA738, IA767, IB827\\
                VISTA & Y, J, H, Ks \\
                \textit{Spitzer} (IRAC) & 3.6~$\mu$m, 4.5~$\mu$m, 5.8~$\mu$m, 8.0~$\mu$m\\
                \quad \quad \quad (MIPS) & 24~$\mu$m\\
                \textit{Herschel} (PACS) & 100~$\mu$m, 160~$\mu$m\\
                \hline
                \end{tabular}
        \label{table:bands}
        \end{center}
\end{table}

\section{Methodology} \label{sec:method}
\subsection{Tools}

\subsubsection{CIGALE} \label{subsec:cigale}
Code Investigating GALaxy Emission \citep[CIGALE; ][]{2009AA...507.1793N} is a spectral energy distribution (SED) modelling and fitting tool with an improved fitting procedure by \citet{2011ApJ...740...22S}. Here, the Python version 0.11.0 is used \citep[][Boquien et al., in prep.]{2005MNRAS.360.1413B, 2009AA...507.1793N} to generate SEDs and to fit these SEDs to the UV to PACS data from COSMOS2015 to estimate the SPIRE 250~$\mu$m, 350~$\mu$m, and 500~$\mu$m flux densities for use as a flux density prior in XID+ (see Sects. \ref{subsec:xidp} and \ref{subsec:scipip}). After the SPIRE band flux densities had been extracted, CIGALE was also used to calculate the physical parameters of each object, such as SFR and M$_{\star}$ as well as rest frame colours. CIGALE uses the energy balance between the attenuated UV emission by dust and the IR emission, allowing the estimation of the FIR flux densities. The reported values and errors for the SPIRE flux densities and physical parameters are created using CIGALE's Bayesian probability density function analysis. CIGALE also gives the flux densities and physical parameters of the best fitting model for each object, but these are not used in this work. 

The choices for the SED model components and parameters for the SPIRE band priors follow \citet{2017A&A...603A.102P}, with a different dust attenuation model and other minor changes, and will briefly be repeated here. We use a delayed exponentially declining star formation history (SFH) with an exponentially declining burst, \citet{2003MNRAS.344.1000B} stellar emission, \citet{2003PASP..115..763C} initial mass function (IMF), \citet{2000ApJ...539..718C} dust attenuation, the updated \citet{2014ApJ...780..172D} version of the \citet{2007ApJ...657..810D} IR dust emission, and \citet{2006MNRAS.366..767F} AGN models. The dust attenuation model was changed from \citet{2000ApJ...533..682C}, used in \citet{2017A&A...603A.102P}, to \citet{2000ApJ...539..718C} as recent work by \citet{2017MNRAS.472.1372L} has shown that the \citet{2000ApJ...533..682C} dust model cannot accurately reproduce the attenuation seen in a sample of dusty galaxies at $z \approx 2$. A list of parameters used, where they differ from the default values, along with a justification can be found in Appendix \ref{app:parameters}.

\subsubsection{XID+} \label{subsec:xidp}
XID+\footnote{\url{https://github.com/H-E-L-P/XID_plus}} \citep{2017MNRAS.464..885H} is a probabilistic de-blending tool used to extract source flux densities from photometry maps that suffer from source confusion. This is achieved by using Bayesian inference to explore the posterior. Once converged, the flux density is reported along with the upper and lower $1\sigma$ uncertainties. In its standard form, XID+ uses a flat prior in parameter space, between zero and the brightest value in the map, along with the source positions on the sky.

This work follows \citet{2017A&A...603A.102P} by using a more informed Gaussian prior, again truncated between zero and the brightest value in the map. The mean and sigma for these Gaussian priors are generated by using CIGALE models to estimate the flux densities for the mean and using two times the error on these estimates as the sigma. To allow parallelisation, which reduces the time taken for XID+ to de-blend the map, the map is split up into tiles based on the Hierarchical Equal Area isoLatitude Pixelization of a sphere system \citep[HEALPix;][]{2005ApJ...622..759G} using order 11, which corresponds to an area of 2.95 arcmin$^{2}$ per tile. Order 11 was chosen as it is a compromise between the number of objects in a tile (more objects means a more reliable flux density extraction) and the time it takes a tile to run.

\subsection{Science pipeline} \label{subsec:scipip}
\begin{figure}
        \centering
        \includegraphics[width=0.5\textwidth]{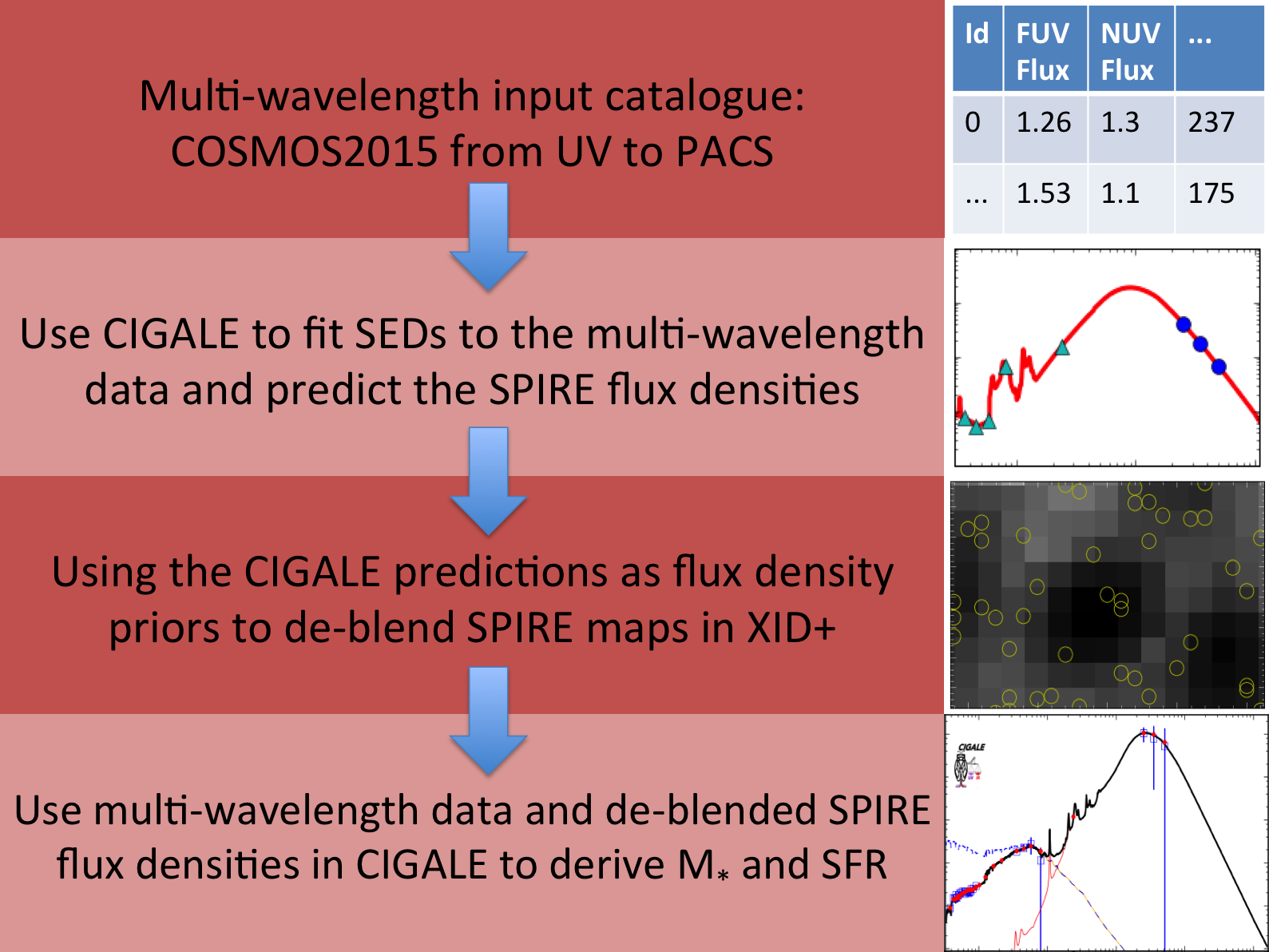}
        \caption{Brief summary of the science pipeline.}
        \label{fig:scipip}
\end{figure}

The  extraction of the flux densities in the SPIRE bands, which follows a similar pipeline to the one used in \citet{2017A&A...603A.102P}, is briefly summarised in Fig. \ref{fig:scipip} and is repeated here for completeness. We begin by using the far-UV (FUV) to PACS 160~$\mu$m data from COSMOS2015 to generate estimates for the flux densities in all three SPIRE bands simultaneously using CIGALE's Bayesian analysis. All objects that were not classified as galaxies (TYPE flag in COSMOS2015 not set to 0, approximately 1.3\%) were removed, as were objects without any photometric redshift (ZPDF $\leq$ 0, approximately 2.9\%). All predicted flux densities were then used in XID+ to extract the flux densities for the objects; we did not remove any objects with poor $\chi^{2}$ values. Once the SPIRE flux densities were extracted, these SPIRE data were added to the FUV-PACS data and CIGALE was rerun, this time to get estimates for M$_{\star}$ and SFR. The same CIGALE models were used for the flux estimation and to obtain the physical parameters so the results from each CIGALE run should not be degenerate.

\begin{figure}
        \centering
        \includegraphics[width=0.45\textwidth]{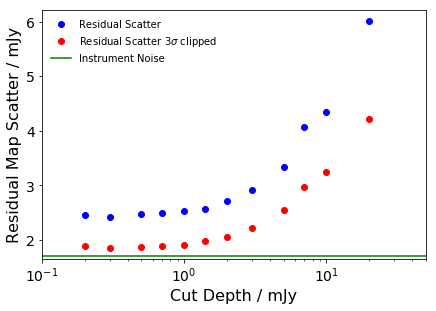}
        \caption{Scatter of the 250~$\mu$m residual map using different depth cuts on the prior list. The blue and red points are with and without 3$\sigma$ clipping, respectively, while the green line is the $1\sigma$ instrument noise for the COSMOS field.}
        \label{fig:resid-scatter}
\end{figure}

The first CIGALE run provides flux density estimates for all the objects in the catalogue. However, the flux density estimates for the faintest objects will be highly uncertain and hence result in extracted flux densities that are unreliable. To find the depth to which we can reliably run XID+, a number of different cut depths on the predicted flux densities at 250~$\mu$m  were used, from 0.2~mJy to 20~mJy, and XID+ run on 25 tiles. A residual map was created by subtracting the replicated map from XID+ from the original image for each depth. In the ideal situation where all sources are correctly accounted for, the residual map should have a scatter consistent with $1\sigma$ instrument noise. Figure \ref{fig:resid-scatter} shows how the scatter of the residual 250~$\mu$m map changes with cut depth with and without 3$\sigma$ clipping. As can be seen, as the cut gets deeper, from 20~mJy to 1~mJy, the scatter reduces;  cut depths deeper than 1~mJy do not cause much of a change. All cut depths have a scatter greater than the instrument noise at 250~$\mu$m (1.71~mJy). This is to be expected as our source list will miss the faintest sources and the flux densities assigned to the known sources will not exactly coincide with the true flux densities for all sources. As a result of all the deeper cuts having a scatter larger than the $1\sigma$ instrument noise, and wanting to keep as large a sample as possible while not risking creating too many degeneracies, a predicted 250~$\mu$m cut of 0.7~mJy was used. For this cut, we ignore any uncertainty on the flux density estimate. This leaves 205\,958
 objects to be run through XID+.

To prevent an overly restrictive prior in XID+ and conservatively capture the uncertainty in SED modelling, the errors on the flux density estimates are expanded by a factor of two (see also Appendix \ref{app:XID+prior}). The estimates of all three SPIRE bands from CIGALE are then used as the priors for XID+. The flux density estimates are used as the means in the XID+ priors while the expanded errors are used as the standard deviations. XID+ was then run on the SPIRE images.
The next step is to run the data from COSMOS2015 and XID+ through CIGALE to generate the required parameters for each object: M$_{\star}$, SFR, and U--V and V--J colours; the colours are needed for the QG/SFG separation.

\subsection{Mass completeness and redshift binning}
To study the MS, we removed all objects with a redshift below 0.2 as the mean error on the photometric redshifts for objects below $z$ = 0.2 is approximately 0.2 (for discussion on the redshifts and their errors, see \citealt{2016ApJS..224...24L}). The remaining objects were then binned by redshift. Each bin's width was determined by using the average error on the redshift of the objects within a bin; the bin size was the average error of all the objects in the bin rounded up to the next  decimal place. Using a bin size of twice the error was also briefly explored and provided consistent results.

\begin{figure}
        \centering
        \includegraphics[width=0.45\textwidth]{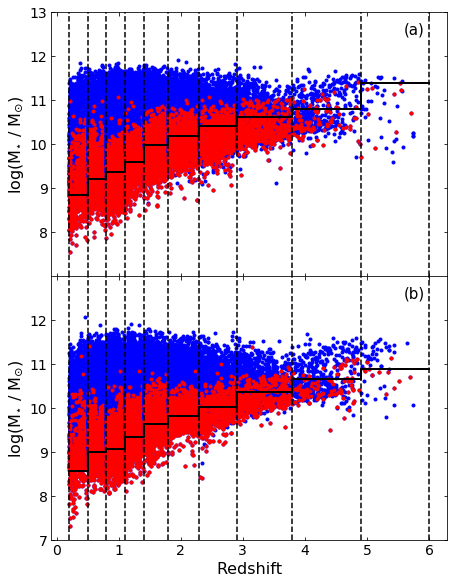}
        \caption{Masses of all objects detected in the Ks band (blue) are shown against redshift along with the faintest 20\% in each redshift bin (red) for the (a) deep and (b) ultra-deep  regions. The 90\% completeness limit is shown by the thick black lines, while the dashed black lines show the edges of each redshift bin.}
        \label{fig:mass_cmplt}
\end{figure}

Once binned by redshift, the galaxies were cut for completeness. This completeness cut was done empirically by following \citet{2010A&A...523A..13P}. The K$_{\mathrm{s}}$ band magnitude limits (K$_{\mathrm{s~lim}}$) were set to 24.7, the 3$\sigma$ limit for the objects within the UltraVISTA \citep{2012A&A...544A.156M} ultra-deep stripes, and 24.0, the 3$\sigma$ limit for the rest of the COSMOS field, in the ultra-deep and deep regions of COSMOS, respectively \citep{2016ApJS..224...24L}. For each galaxy with a redshift (ZPDF > 0) and a K$_{\mathrm{s}}$ band detection the mass the galaxy would need (M$_{\mathrm{lim}}$) to be observed at the magnitude limit was calculated using
\begin{equation}\label{eq:mass}
        \log(\mathrm{M}_{\mathrm{lim}}) = \log(\mathrm{M}) - 0.4(\mathrm{K}_{\mathrm{s~lim}} - \mathrm{K}_{\mathrm{s}}),
\end{equation}
where M is the galaxy's mass in M$_{\odot}$ and K$_{\mathrm{s}}$ is the galaxy's K$_{\mathrm{s}}$ band magnitude. In each redshift bin, the faintest 20\% of objects were selected and the limiting mass was the M$_{\mathrm{lim}}$ value which 90\% of these faintest objects lie below. As not all objects have a K$_{s}$ band detection (the catalogue is zYJHKs selected), these limiting mass data points were used for all objects. The mass limits for the deep and ultra-deep regions of COSMOS are shown in Fig \ref{fig:mass_cmplt}.

\subsection{Forward modelling}
To determine the MS in each redshift bin, we use the Markov chain Monte Carlo (MCMC) sampler emcee \citep{2013PASP..125..306F} to sample the parameter space of the chosen MS model. For our routine, we create model SFRs using the observed M$_{\star}$, observed redshift, and the MS being tested at each step. For each M$_{\star}$, a random number is drawn from a Gaussian distribution centred on the SFR of the MS at that M$_{\star}$ and with the scatter of the MS as the standard deviation. The Gaussian distribution is also truncated such that it reproduces the observed upper and lower SFR limit. These upper and lower SFR limits are determined by finding the SFR in each redshift bin that 0.1\% of the objects fall above or below, and then fitting to these values as a function of redshift using 
\begin{equation}
        S_{lim}(z) = [B_{0} \times \log(B_{1}+z)] + B_{2},
\end{equation}
where $S$ is $\log(\mathrm{SFR}/\mathrm{M}_{\odot} \mathrm{yr}^{-1})$ and $B_{n}$ are the coefficients that are found. The parameters $B_0$,  $B_1$, and $B_2$, take the values 0.61 $\pm$ 0.32, -0.37 $\pm$ 0.01~$\mathrm{M}_{\odot} \mathrm{yr}^{-1}$,  and 3.13 $\pm$ 0.13~$\log(\mathrm{M}_{\odot} \mathrm{yr}^{-1})$ for the upper limit and 2.86 $\pm$ 0.19, -0.17 $\pm$ 0.05~$\mathrm{M}_{\odot} \mathrm{yr}^{-1}$, and 0.12 $\pm$ 0.10~$\log(\mathrm{M}_{\odot} \mathrm{yr}^{-1})$ for the lower limit, respectively. The upper and lower limits are applied to each simulated object individually using the observed redshifts.

\begin{figure}
        \centering
        \includegraphics[width=0.5\textwidth]{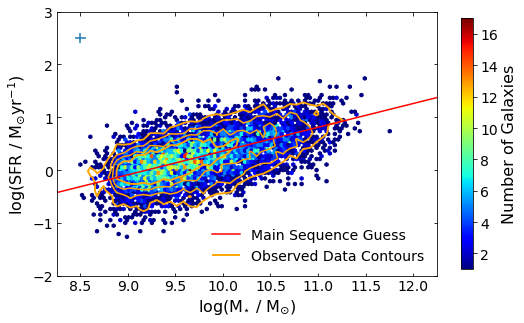}
        \caption{Example of the data generated at one step of the MCMC routine for the lowest ($0.2 \leq z < 0.5$) redshift bin, shown as number density from high (dark red) to low (dark blue). The MS being tested at this step, in this case the most likely step, is shown as the red line, while the contours for number density of the observed data are shown in orange. The size of the average observed error on SFR and M$_{\star}$ is also shown as a blue cross. }
        \label{fig:ex-step}
\end{figure}

Once the model SFRs have been generated, both the SFR and M$_{\star}$ are perturbed by adding a second random number drawn from a Gaussian centred on zero and with a standard deviation equal to the error on the observed SFR or M$_{\star}$ of that individual object. An example of the MS generated at one step, assuming a linear power law, is shown in Fig. \ref{fig:ex-step}, along with the MS used to generate at that step and the contours of the observed data.

To compare the model data to the observed data at each step, the two data sets are binned by stellar mass into identical bins with a width of 0.25~dex. The mean and standard deviation of the SFRs in each mass bin are calculated and the mean and standard deviations of the model data are compared to their counterparts from the observed data. The smaller the differences, the greater the likelihood that the model is a correct representation of the observed data.

\begin{figure}
        \centering
        \includegraphics[width=0.5\textwidth]{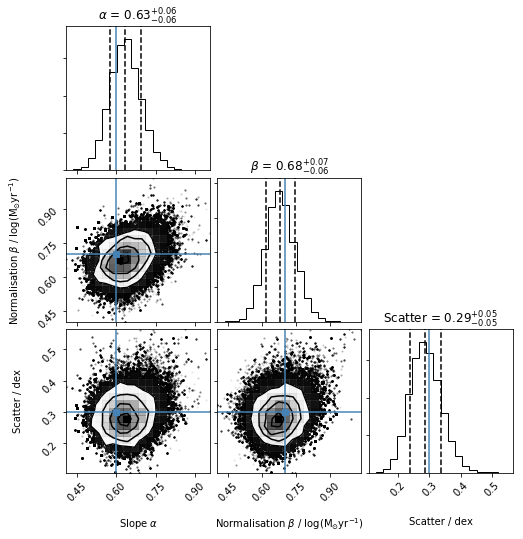}
        \caption{Corner plot \citep{corner} of the marginalised posterior of the forward modelling routine applied to a mock data set with known slope (0.6), normalisation (0.7 $\log(\mathrm{M}_{\odot} / \mathrm{yr}^{-1})$), and scatter (0.3 dex), shown as the blue lines. Panels on the diagonal show the 1D marginalised posteriors for the slope, normalisation and scatter (left to right). Off-diagonal panels show the combined 2D posteriors as labelled by their axes. The recovered 16th, 50th, and 84th percentiles are shown by the dashed vertical lines; all the input parameters are recovered, within error.}
        \label{fig:fm_example}
\end{figure}

All the parameters in the models used were treated as if they were uncorrelated, although this is not strictly true. However, this method was found to accurately recover input relations used to generate mock sets of data that assume a linear power law. Figure \ref{fig:fm_example} shows an example of the posterior for a mock data set with known slope (0.6), normalisation (0.7 $\log(\mathrm{M}_{\odot} / \mathrm{yr}^{-1})$), and scatter (0.3 dex). As can be seen, the input parameters are recovered within   error.

\section{Results} \label{sec:results}
The M$_{\star}$ found using CIGALE are compared to those found by \citet{2016ApJS..224...24L} in COSMOS2015. We find that our M$_{\star}$ are higher, on average, by 0.15~dex and are consistent with COSMOS2015 within the COSMOS2015 average error of 0.15~dex, but not the 0.10~dex average error of our M$_{\star}$. Both \citet{2016ApJS..224...24L} and this work use the \citet{2003PASP..115..763C} IMF and the \citet{2003MNRAS.344.1000B} stellar population model, so this is not the cause of the slight discrepancy, but the dust attenuation models differ: this work uses the \citet{2000ApJ...539..718C} dust attenuation, while \citet{2016ApJS..224...24L} uses the \citet{2000ApJ...533..682C} dust attenuation. Recently, \citet{2017MNRAS.472.1372L} have investigated the effects of dust attenuation laws on the physical properties of dusty galaxies at $z \approx 2$. They find that using the \citet{2000ApJ...539..718C} dust attenuation results in higher M$_{\star}$ values than \citet{2000ApJ...533..682C}, resulting in M$_{\star}$ higher by a factor of 1.4, or approximately 0.15~dex. This is consistent with the difference in M$_{\star}$ found between our current work and \citet{2016ApJS..224...24L}. Thus, we conclude that the difference in M$_{\star}$ is a result of the choice of dust attenuation model.

To remove the QG, we follow the \citet{2011ApJ...735...86W} UVJ colour cut
\begin{equation}\label{eqn:UVJ}
        \begin{aligned}
                (U - V) > 0.88 \times (V - J) + 0.69 && z < 0.5\\
                (U - V) > 0.88 \times (V - J) + 0.59 && z > 0.5\\
                (U - V) > 1.3, (V - J) < 1.6 && z < 1.5\\
                (U - V) > 1.3, (V - J) < 1.5 && 1.5 < z < 2.0\\
                (U - V) > 1.2, (V - J) < 1.4 && 2.0 < z < 3.5\\
        \end{aligned}
\end{equation}
where the rest-frame (U -- V) and  (V -- J)  colours come from the second CIGALE run. If these conditions are not met, the object is SF. The (U -- V) and (V -- J) criteria for $2.0 < z < 3.5$ were expanded for all objects with a redshift greater than 2, such that the final line in Eq. \ref{eqn:UVJ} becomes
\begin{equation}\label{eqn:UVJ-new}
        \begin{aligned}
                (U - V) > 1.2, (V - J) < 1.4 && 2.0 < z < 6.0.\\
        \end{aligned}
\end{equation}

Once the QG objects are removed, we fit two models to the data: the \citet{2015ApJ...801...80L} description,  which contains a turn-over
\begin{equation}\label{eqn:lee}
        S = S_{0} - \mathrm{log}\Bigg[ 1 + \Bigg(\frac{\mathrm{M}_{\star}}{\mathrm{M}_{0}}\Bigg)^{-\gamma}\Bigg],
\end{equation}
where $S$ is $\log(\mathrm{SFR}/\mathrm{M}_{\odot} \mathrm{yr}^{-1})$, $S_{0}$ is the maximum value of $S$ that the function approaches at high mass, M$_{0}$ is the turn-over mass in M$_{\odot}$, and $\gamma$ is the low-mass slope, and the \citet{2012ApJ...754L..29W} single power law
\begin{equation}\label{eqn:whitaker}
        S = \alpha[\log(\mathrm{M}_{\star}) - 10.5] + \beta,
\end{equation}
where $\alpha$ is the slope and $\beta$ is the normalisation at $\log(\mathrm{M}_{\star}/\mathrm{M}_{\odot}) = 10.5$.

\begin{figure}
        \centering
        \includegraphics[width=0.5\textwidth]{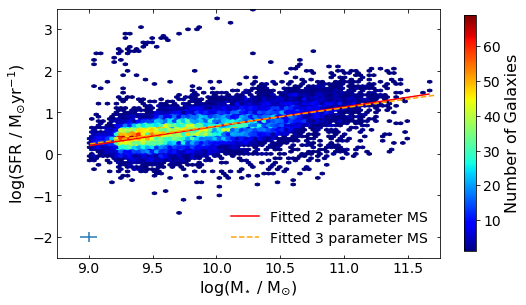}
        \caption{Comparison of fitting with Eq. \ref{eqn:lee} (orange dashed line) and Eq. \ref{eqn:whitaker} (red line) in the $0.5 \leq z < 0.8$ redshift bin. The galaxies are shown as a number density plot, with dark red being high number density and dark blue low number density, and the size of the average error on SFR and M$_{\star}$ is shown as a blue cross. There is very little difference in shape between the two fits, demonstrating that Eq. \ref{eqn:whitaker} is the preferred form of the MS.}
        \label{fig:3param}
\end{figure}

Fitting with Eq. \ref{eqn:lee} was reasonably unsuccessful. No redshift bins show significant evidence of a high-mass turn-over with all redshift bins being consistent with a simple power law, as can be seen in the example in Fig. \ref{fig:3param}. The majority of the turn-over masses found from the MCMC forward modelling are larger than the highest mass object within each redshift bin, approximately $\log(\mathrm{M}_{\star} / \mathrm{M}_{\odot}) = 12.5$. As a result, these cannot be considered  reliable results as the turn-over position is unconstrained by the data. The turn-over positions at all redshifts, assuming they are reliable, are also much higher than those found in the literature \citep[$\log(\mathrm{M}_{0} / \mathrm{M}_{\odot}) \approx 10.5$ e.g.][]{2015ApJ...801...80L, 2016ApJ...817..118T}.  We also found that when fitting Eq. \ref{eqn:lee} with $\gamma$ held fixed at 1.17, which was shown by \citet{2015ApJ...801...80L} to be reasonable and also leaves the same number of free parameters as Eq. \ref{eqn:whitaker}, the majority of the redshift bins have a best fit that is less probable than the best fit using Eq. \ref{eqn:whitaker}. As a result, we conclude that Eq. \ref{eqn:whitaker} is a better description of our data.

\begin{figure*}
        \centering
        \includegraphics[height=0.919\textheight]{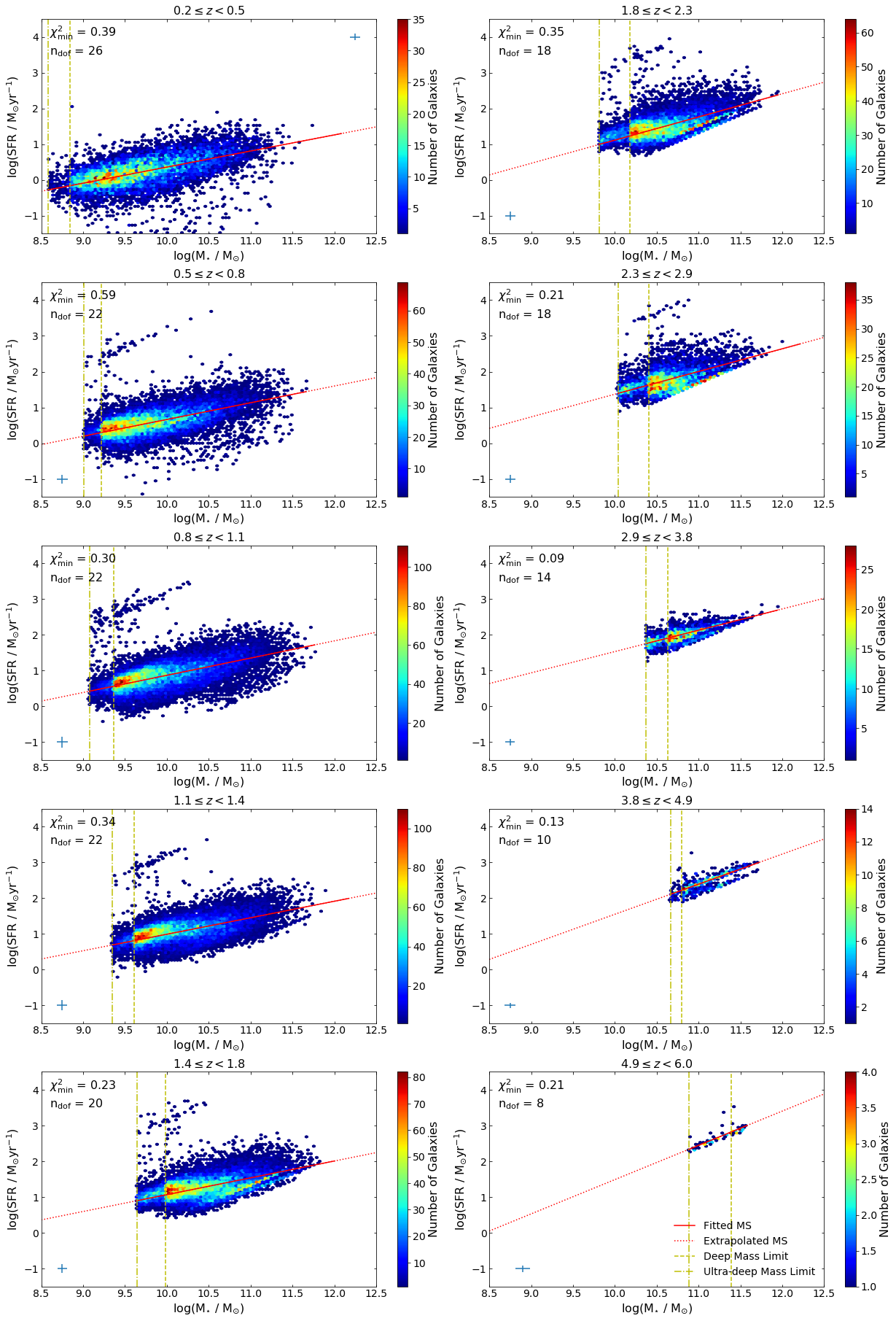}
        \caption{Fitting of Eq. \ref{eqn:whitaker} to the objects in the redshift bins as labelled. The solid line is the most likely MS across the fitted M$_{\star}$ range, while the dotted line is an extrapolation across the M$_{\star}$ range of the plot. The galaxies are shown as a number density plot, with dark red being high number density and dark blue low number density. The vertical density discontinuities are the result of the two depths of data used: the deep mass limit is the yellow dashed line and the ultra-deep mass limit is the dot-dashed yellow line. Each panel also indicates the $\chi^{2}$ of the most likely MS ($\chi^{2}_{\mathrm{min}}$) and the number of degrees of freedom in the fitting (n$_{\mathrm{dof}}$), and shows the size of the average SFR and M$_{\star}$ errors as a blue cross.}
        \label{fig:exMS}
\end{figure*}

\begin{figure}
        \centering
        \includegraphics[width=0.5\textwidth]{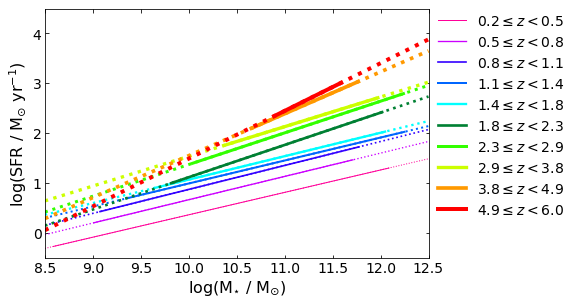}
        \caption{Fitted MS trends using Eq. \ref{eqn:whitaker}. The solid and dashed lines are the MS across the fitted M$_{\star}$ range. The dotted lines are extrapolations across the M$_{\star}$ range of the plot. The normalisation of the MS clearly increases as redshift increases. The slight decrease in slope at low redshift can be seen along with the increase above $z = 1.1$. The very steep slopes at $1.8 \leq z < 2.9$ can also be seen.}
        \label{fig:UVJ-MS}
\end{figure}

\begin{figure}
        \centering
        \includegraphics[width=0.5\textwidth]{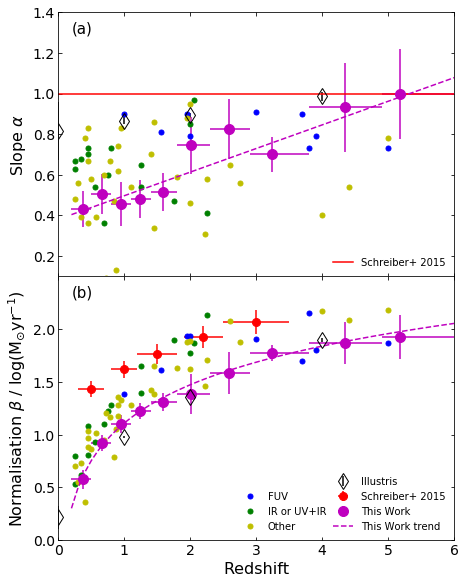}
        \caption{Comparison of the UVJ selected MS results of this work (magenta) with the observational MS from \citet[][FUV data in blue; IR data in green; and radio, hydrogen lines, and UV SED fitting in yellow]{2014ApJS..214...15S}, \citet{2015A&A...575A..74S} low mass MS (red), and the MS from the Illustris Simulation \citep[][black diamonds]{2014MNRAS.444.1518V, 2015MNRAS.447.3548S}. The $\alpha$ and $\beta$ parameters from Eq. \ref{eqn:whitaker} are in panels (a) and (b), respectively. As \citet{2015A&A...575A..74S} hold their low-mass slope constant at unity, this is indicated in panel (a) as a flat red line. The redshifts shown for this work are the mean redshift in each redshift bin, while the horizontal error bars show the width of the redshift bin. A version of this plot using SFRs derived from IR template fitting can be found in Appendix \ref{app:schreiber}.}
        \label{fig:work-comp}
\end{figure}

With the knowledge that our data do not support a turn-over in the MS, Eq. \ref{eqn:whitaker} was also fitted to our data, as shown in Fig. \ref{fig:exMS}. This form of the MS was much more successful, with well-constrained fitting parameters in all redshift bins (see Table \ref{table:whitaker-params} and Fig. \ref{fig:UVJ-MS}). The $\beta$ parameter (normalisation) increases rapidly with redshift out to $z \approx 2$ before increasing more slowly (see magenta points in Fig. \ref{fig:work-comp}b). The evolution of $\beta$ was found, fitting using the SciPy \citep{Scipy} orthogonal distance regression \citep[ODR;][]{MR1087109} package, to follow
\begin{equation}\label{eqn:beta-UVJ}
        \beta(z) = (1.10 \pm 0.07) + [(0.53 \pm 0.05) \times \ln(\{0.03 \pm 0.11\} + z)],
\end{equation}
with errors in the coefficients from the ODR fitting.

The $\alpha$ parameter (slope) is slightly less smooth in its evolution (see Fig. \ref{fig:work-comp}a, magenta). Generally, the slope increases with redshift across the entire redshift range of this study. There is a potential rise between $z \approx 1.8$ and $z \approx 2.9$, although this is very marginal and may be a result of the SED model gridding causing a higher slope (see Fig. \ref{fig:exMS}). A linear relation was used to find the slope evolution, which was determined to be
\begin{equation}\label{eqn:alpha-UVJ}
        \alpha(z) = (0.38 \pm 0.04) + (0.12 \pm 0.02)z.
\end{equation}

Short tests were also done using just the deep and ultra-deep data. At redshifts below 4.9, just using the deep or ultra-deep data made no significant difference to the $\alpha$ and $\beta$ parameters: the values change within  error, compared to using both the deep and ultra-deep data together. With just the deep data, the slope of the highest redshift bin becomes much lower, although it is consistent within  error. For normalisation, the deep data has a lower normalisation, but again it is consistent within  error. This is likely a result of low number statistics. There are only 16 UVJ selected objects with $z \geq 4.9$ in the deep coverage. With so few objects to fit with, the results are highly uncertain with just the deep data.

\begin{table*}\label{table:whitaker-params}
        \caption{Parameters from fitting Eq. \ref{eqn:whitaker} to the UVJ selected star forming galaxies, with $\alpha$  the slope and $\beta$  the normalisation at $\log$(M$_{\star}$/M$_{\odot}$) = 10.5. The intrinsic scatter is found during the MCMC fitting.}
        \centering
        \begin{tabular}{l c c c c c c c c}
                \hline
                Redshift Bin & Average Redshift & Number of Sources & $\alpha$ & $\alpha$ Error & $\beta$ & $\beta$ Error & Scatter & Scatter Error\\
                & & & & & $\log(\mathrm{M}_{\odot} \mathrm{yr}^{-1})$& $\log(\mathrm{M}_{\odot} \mathrm{yr}^{-1})$ & dex & dex\\
                \hline
                0.2$\leq z<$0.5 & 0.37 & 11460 & 0.43 & 0.09 & 0.58 & 0.09 & 0.35 & 0.06\\
                0.5$\leq z<$0.8 & 0.66 & 17615 & 0.50 & 0.10 & 0.92 & 0.08 & 0.33 & 0.06\\
                0.8$\leq z<$1.1 & 0.95 & 24537 & 0.46 & 0.11 & 1.10 & 0.08 & 0.34 & 0.06\\
                1.1$\leq z<$1.4 & 1.24 & 20712 & 0.48 & 0.09 & 1.22 & 0.07 & 0.28 & 0.06\\
                1.4$\leq z<$1.8 & 1.59 & 19388 & 0.51 & 0.09 & 1.31 & 0.08 & 0.24 & 0.05\\
                1.8$\leq z<$2.3 & 2.02 & 13166 & 0.74 & 0.14 & 1.39 & 0.19 & 0.29 & 0.08\\
                2.3$\leq z<$2.9 & 2.59 & 6375 & 0.83 & 0.15 & 1.59 & 0.20 & 0.28 & 0.07\\
                2.9$\leq z<$3.8 & 3.23 & 2338 & 0.70 & 0.09 & 1.77 & 0.08 & 0.10 & 0.05\\
                3.8$\leq z<$4.9 & 4.34 & 590 & 0.93 & 0.22 & 1.87 & 0.20 & 0.15 & 0.07\\
                4.9$\leq z<$6.0 & 5.18 & 72 & 1.00 & 0.22 & 1.92 & 0.21 & 0.08 & 0.05\\
                \hline
        \end{tabular}
\end{table*}

\begin{figure}
        \centering
        \includegraphics[width=0.45\textwidth]{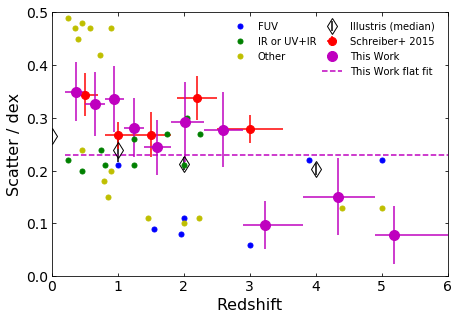}
        \caption{Intrinsic scatter about the MS found from the MCMC fitting of the data in each redshift bin to Eq. \ref{eqn:whitaker}. This work (magenta) is shown along with the observed scatters from \citet[][FUV data in blue, IR data in green, and other data in yellow]{2014ApJS..214...15S} and \citet[][red]{2015A&A...575A..74S} as well as the median scatter found at each redshift in the Illustris Simulation \citep[][black diamonds]{2014MNRAS.445..175G, 2014MNRAS.444.1518V, 2015MNRAS.447.3548S}. The redshifts shown for this work are the mean redshift in each redshift bin while the horizontal error bars show the width of the redshift bin. Also shown is the best fit to this work's intrinsic scatter assuming no redshift evolution (dashed magenta line). The intrinsic scatter found in this work is consistent with existing literature, although above $z \approx 1.8$ our intrinsic scatter is smaller than is found in works that use IR SFR traces.}
        \label{fig:scatter}
\end{figure}

After determining the MS, we check the scatter to ensure that we have actually found the MS. The intrinsic scatter was determined during the MCMC fitting and was assumed to be independent of M$_{\star}$, which is believed to be true \citep{2012ApJ...754L..29W, 2014ApJS..214...15S, 2016ApJ...817..118T}. Our intrinsic scatter is found to be consistent with other literature \citep[e.g.][see Fig. \ref{fig:scatter}]{2012ApJ...754L..29W, 2014ApJS..214...15S, 2015MNRAS.447.3548S, 2015A&A...575A..74S, 2016ApJ...817..118T}, suggesting that we have indeed found the MS. We also fit a flat line to our intrinsic scatter, assuming that there is no redshift evolution, which gives an amplitude of $0.23 \pm 0.03$~dex, lower than the expected 0.3~dex typically found in observational works but consistent with the results from the Illustris Simulation \citep{2014MNRAS.445..175G, 2014MNRAS.444.1518V, 2015MNRAS.447.3548S}. Our small intrinsic scatter may be an underestimation resulting from the M$_{\star}$ and SFR being derived from the same SED fitting procedure. This results in some correlation between these two quantities and hence reduces the scatter about the MS.

\section{Discussion}\label{sec:discussion}
\subsection{Form of the main sequence}
In this work, we find little evidence of a turn-over. Thus, any cause of the reduction in specific SFR (SFR/M$_{\star}$, sSFR) with mass does not suddenly `turn on', rather it is an effect that  slowly increases with M$_{\star}$. An increasing slope with redshift indicates that the mass dependent reduction in sSFR becomes weaker as we look further back, meaning that the mechanism behind this was less strong in the past than it is now.

Second, at relatively low redshift we find that the slope is relatively shallow, which suggests that the ability of galaxies to form stars reduces as mass increases. This may be a result of a decrease in star formation efficiency, a reduction in the available cold gas, or some other mass-dependent quenching mechanism. We also see that the slope increases with redshift, implying that galaxies in the younger universe were more self-similar than they are today. Thus, in the modern universe, large galaxies have lower specific SFRs (sSFR) than smaller galaxies, while at high redshift, the sSFRs of large and small galaxies are similar.

Third, we find that the slope of the MS is very high between redshifts of 1.8 and 2.9 with respect to other epochs. We do note, however, that the significance of this rise is very marginal. A steep slope at these times suggests a higher sSFR in large galaxies compared to other times. This increase in the slope coincides with the peak of the cosmic star formation history (CSFH) at $z \approx 2-3$ \citep{2014ARA&A..52..415M}. It may also simply be a result of the SED model gridding causing a higher slope, indicated by the discontinuities in the two top right panels of Fig. \ref{fig:exMS}.

When the normalisation is also considered, the low-mass end of the MS also sees changes in a galaxy's ability to form stars. Equation \ref{eqn:whitaker} has a normalisation at $\log(\mathrm{M}_{\star} / \mathrm{M}_{\odot}) = 10.5$ and it does not evolve fast enough to maintain a constant crossing point of the MS at all redshifts; the intercept mass of the MS in one redshift bin with the MS in the lowest redshift bin increases out to $z \approx 1.8$ before becoming approximately constant, as can be seen in Fig. \ref{fig:UVJ-MS}. Thus, the low-mass galaxies must have a decrease in sSFR as redshift increases, out to $z \approx 1.8$, while the high-mass galaxies have an increasing sSFR.

\subsection{Contamination by quiescent galaxies}\label{subsec:z}
We find the main sequences in the highest redshift bins ($z > 1.8$) are identical regardless of whether we cut for SFG or not. This may be a result of the dusty SFG and QG at high redshift having their near-IR (NIR) observed frame emission suppressed below the detection limits of the NIR selected catalogue by dust obscuration in the galaxy. Alternatively, there is a greater mixing of SFG and QG at lower redshifts and higher masses, pulling down the MS when the QG classed objects are removed. The latter is likely the correct scenario. There is little evidence for a turn-over at lower redshifts after selection of the SFGs. We thus surmise that the turn-over in other works may be the result of a SFG selection that is not effective enough in removing the QG. This would result in an  overabundance of low SFR galaxies at higher mass which act to pull the high-mass  MS down, similar to the conclusions of \citet{2015MNRAS.453.2540J}. It is the lack of turn-off in the MS at lower redshifts that leads us to believe that the high-$z$ colour cut employed (Eq. \ref{eqn:UVJ}) can be extended to all redshifts above $z = 2.0$ as we also recover the linear MS at these higher redshifts.

\subsection{Comparisons with other works}\label{subsec:otherwork}
The work by \citet{2014ApJS..214...15S} collates a large number of MS studies and generates a consensus regarding the MS. This provides an excellent base line to which we can compare this work. As with the consensus result, this work finds little evidence of a high-mass turn-over. Broadly, our results agree with \citet{2014ApJS..214...15S}: both the slope and normalisation increase with redshift;  the rate of change in normalisation decreases as redshift increases. However, quantitatively there are differences. 

The normalisation found in this work is in reasonable agreement with the \citet{2014ApJS..214...15S} collection (see Fig. \ref{fig:work-comp}b), but lies towards the bottom of the collection. This slight lowering of the normalisation, especially compared with studies using IR SFR indicators, may be due to the splitting of brighter objects into a number of fainter sources.

The slope of this work is consistent with the \citet{2014ApJS..214...15S} results at all redshifts, rising from the bottom of the collection at low redshifts to the top of the collection at high redshifts. Between $z \approx 2$ and $z \approx 3$ we are approximately consistent with the far-ultraviolet  SFR indicated MS, but are typically higher than the MS using other (radio, hydrogen lines and UV SED fitting) SFR indicators. This is potentially a result of the data in this work looking through the dust with the \textit{Herschel} data, which will raise the SFR of the high-mass, dusty galaxies. However, this should also result in the FUV MS having a shallower slope than this work as well.

The work by \citet{2015A&A...575A..74S} looked at the MS using Herschel, but their study relied on stacking rather than de-blending. We convert the \citet{2015A&A...575A..74S} results from the \citet{1955ApJ...121..161S} IMF to the \citet{2003PASP..115..763C} IMF used here, following \citet[see Eq. 2]{2014ApJS..214...15S} to correct the mass and \citet{2012ARA&A..50..531K} to correct the SFR. Their results found a turn-over at high mass, unlike the results of this study. This high-mass turn-over is not a result of stacking;  we performed our own stacking analysis and found no turn-over or UVJ selection because our UVJ selection is similar to that of \citet{2015A&A...575A..74S} (see Appendix \ref{app:schreiber}).

As a result of their turn-over, just the low-mass slope of \citet{2015A&A...575A..74S} is used for comparison in  this work. \citet{2015A&A...575A..74S} also enforce a slope of unity while the normalisation is allowed to change. A non-redshift dependant slope of unity is not found to be a good description of our data, with the slope remaining below this until the highest redshift bin. We also find that our normalisation is considerably lower at all redshifts, by approximately 0.4~dex, with the difference being larger at lower redshifts (see Fig. \ref{fig:work-comp}). However, this difference in normalisation is a result of the different methods used to derive the SFR of the galaxies, with our SED-based SFR producing values lower by 0.4~dex when compared to the SFRs derived from the IR luminosity from CIGALE. We also see the same 0.4~dex offset when deriving our own IR luminosity SFRs by fitting our SPIRE data to \citet{2001ApJ...556..562C} templates (see Appendix \ref{app:schreiber}).

We also compare our results to those from the Illustris Simulation \citep{2014MNRAS.445..175G, 2014MNRAS.444.1518V, 2015MNRAS.447.3548S}. We fit Eq. \ref{eqn:whitaker} to the SFR-M$_{\star}$ positions in Fig. 1 of \citet{2015MNRAS.447.3548S} for the $z =$ 0, 1, 2, and 4 redshift bins. The $z = 0$ bin shows clear evidence of a turn-over, but we still fit a simple power law to all the data in this redshift bin. The resulting $\alpha$ and $\beta$ values can be found in Fig. \ref{fig:work-comp} as black diamonds. Our work has slopes that are much shallower than the Illustris values  below a redshift of $\approx 1.8$. At $z \approx 2$, this work and Illustris agree within error, while at $z \approx 4$ our slope appears to be consistent with Illustris. The normalisation of the MS from the Illustris simulation, along with other simulations, is known to be too low with respect to observations, certainly at $z = 1-2$ \citep{2015MNRAS.447.3548S, 2015MNRAS.450.4486F}. Our normalisation results are not consistent with this picture; the normalisation we find is consistent with the results from Illustris. This may suggest that the low normalisations found in simulations may be the result of simulations not being affected by the source blending that observational studies can show.

\section{Conclusions} \label{sec:conc}
The MS is a tight relation between M$_{\star}$ and the SFR of galaxies in the universe. The FIR is a very important part of the spectrum for determining the SFR of an object; however, current cosmological FIR surveys suffer from poor resolution. Here, the latest de-blending tool, XID+, and techniques have been used to break through the \textit{Herschel} confusion limit, allowing better SFR estimates to be generated for nearly a quarter of a million objects in the COSMOS field. This has allowed the examination of the main sequence of star forming galaxies beyond the \textit{Herschel} confusion limit.

The objects were binned by redshift between $0.2 \leq z < 6.0$ and fitted with both a power law and a power law with a high-mass turn-over. The high-mass turn-over form of the MS failed to provide reasonable values for its parameters. The simple power law, on the other hand, produced well-behaved parameters in every redshift bin. Thus, we conclude that the MS using multi-wavelength data and our de-blended \textit{Herschel} SPIRE data shows little evidence for a high-mass turn-over.

We find that the normalisation of the MS increases with redshift, rapidly out to $z \approx 2$ and more slowly thereafter. The slope has a different evolution, rising steadily across all redshifts, with perhaps a weak indication of a brief peak at $1.8 \leq z < 2.9$. An increase in slope with redshift is likely a result of high-mass galaxies having an increase in star formation efficiency with respect to the low-mass galaxies. This results in all galaxies becoming more self similar with increasing redshift, as the slope increases towards unity. The brief peak in slope, if present, appears to coincide with the peak of the CSFH.

Our normalisation results are consistent with the \citet{2014ApJS..214...15S} MS compilation results within error, and with the \citet{2015A&A...575A..74S} stacked \textit{Herschel} MS, once adjusted for IMF and SFR tracers. The slope of our MS is also consistent with the \citet{2014ApJS..214...15S} compilation but moves from the lower regions at low redshift to the upper regions at higher redshifts. This may be a result of hidden star formation being revealed by the \textit{Herschel} data, demonstrating the importance of the FIR/sub-mm for the MS.

\begin{acknowledgements}
We would like to thank the anonymous referee whose comments greatly improved this paper.
We would like to thank the Center for Information Technology of the University of Groningen for their support and for providing access to the Peregrine high performance computing cluster.
This research has made use of data from HerMES project (http://hermes.sussex.ac.uk/). HerMES is a Herschel Key Programme utilising Guaranteed Time from the SPIRE instrument team, ESAC scientists, and a mission scientist.
The HerMES data were accessed through the Herschel Database in Marseille (HeDaM - http://hedam.lam.fr) operated by CeSAM and hosted by the Laboratoire d'Astrophysique de Marseille.
HerMES DR4 was made possible through support of the Herschel Extragalactic Legacy Project, HELP (http://herschel.sussex.ac.uk).
PDH and SJO would like to acknowledge the research leading to these results, which has received funding from the European Union Seventh Framework Programme FP7/2007-2013/ under grant agreement n\degr 607254. This publication reflects only the author's view and the European Union is not responsible for any use that may be made of the information contained therein.
\end{acknowledgements}

\bibliographystyle{aa} 
\bibliography{Paper-AA-2018-32821}

\begin{appendix}
\section{CIGALE Parameters}\label{app:parameters}
In this work we closely follow the CIGALE model selection of \citet[][Appendix A]{2017A&A...603A.102P}. A delayed exponentially declining SFH with an exponentially declining burst was used over the more commonly used exponentially declining or delayed exponential SFH as these two SFHs did not appear to reproduce the expected starburst population in the star formation rate versus stellar mass plane. The e-folding time of the two stellar populations (old and young) in the SFH was roughly matched to that of \citet{2013MNRAS.435...87M}. As \citet{2013MNRAS.435...87M} used a single declining exponential SFH, the e-folding times were split with the burst population taking values of 9~Gyr and above, and the main population taking values of less than 9~Gyr. For the ages of the main population, the values were sampled linearly between 1 and 11~Gyr with a wider sampling up to 13~Gyr. The mass fraction of the burst population follows \citet{2015AA...576A..10C} along with the age of the young stellar population, which also had a lower age of 0.001~Gyr added.  The \citet{2003MNRAS.344.1000B} stellar population model was used with a \citet{2003PASP..115..763C} initial mass function. As this study was not to explore the metallicity of galaxies, it was decided to leave the metallically at solar.

Recent work by \citet{2017MNRAS.472.1372L} has shown that a single power law is a poor representation of the attenuation that occurs in dusty galaxies. As such, we adopt a double power law, similar to that of \citet{2000ApJ...539..718C}, with individual power laws for the birth clouds (BCs) and inter stellar medium (ISM). However, we use a smaller range of values for the V-band attenuation in the BCs than \citet{2000ApJ...539..718C}, a maximum value of 3.8, as tests with higher attenuation values do not noticeably effect the results.

For the dust emission, the polycyclic aromatic hydrocarbon (PAH) fraction had an increase in range around the default 2.5 so more fractions could be sampled while keeping the number of models reasonable. The minimum scaling factor of the radiation field was similarly given a range to sample with an increase in the smallest value from 1.0 to 5.0. The illuminated fraction was reduced to 0.02, following \citet{2015AA...576A..10C}.

The parameters in the AGN model were matched to those used by \citet{2015AA...576A..10C}, who undertook a detailed study of AGN host galaxy emission using CIGALE. The number of choices of frac$_{AGN}$ was reduced from 14 to 5, while still covering the same range, to reduce the number of models created by CIGALE and hence decrease runtime.

A list of parameters, where they differ from default, can be found in Table \ref{table:params}.1.

\begin{table*}\label{table:params}
        \caption{Parameters used for the various properties in the CIGALE model SEDs where they differ from the default values. All ages and times are in Gyr.}
        \label{table:parameters}
        \centering
        \begin{tabular}{l c c}
                \hline
                Parameter & Value & Description\\
                \hline
                \multicolumn{3}{c}{Star Formation History} \\
                \hline
                & & \\
                $\tau_{main}$ & 1.0, 1.8, 3.0, 5.0, 7.0 & e-folding time (main)\\
                $\tau_{burst}$ & 9.0, 13.0 & e-folding time (burst) \\
                $f_{burst}$ & 0.00, 0.10, 0.20, 0.30 & Burst mass fraction\\
                Age & 1.00, 1.50, 2.00, 2.50, 3.00, 3.50, 4.00, & Population age (main)\\
                 & 4.50, 5.00, 5.50, 6.00, 6.50, 7.00, 7.50, & \\
                 & 8.00, 8.50, 9.00, 9.50, 10.00, 10.50, & \\
                 & 11.00, 12.00, 13.00 & \\
                Burst Age & 0.001, 0.010, 0.030, 0.100, 0.300 & Population age (burst)\\
                & & \\
                \hline
                \multicolumn{3}{c}{Stellar Emission}\\
                \hline
                & & \\
                IMF & \citet{2003PASP..115..763C} & Initial Mass Function\\
                $Z$ & 0.02 & Metallicity (0.02 is Solar)\\
                Separation Age & 0.01 & Separation between young and old stellar populations\\
                & & \\
                \hline
                \multicolumn{3}{c}{Dust Attenuation}\\
                \hline
                & & \\
                A$_V^{BC}$ & 0.3, 1.2, 2.3, 3.3, 3.8 & V-band attenuation of the birth clouds\\
                Slope$_{BC}$ & -0.7 & Birth cloud attenuation power law slope\\
                BC to ISM Factor & 0.3, 0.5, 0.8, 1.0 & Ratio of the birth cloud attenuation to ISM attenuation\\
                Slope$_{ISM}$ & -0.7 & ISM attenuation power law slope\\
                & & \\
                \hline
                \multicolumn{3}{c}{Dust Emission}\\
                \hline
                & & \\
                $q_{PAH}$ & 0.47, 1.12, 2.50, 3.9 & Mass fraction of PAH \\
                $U_{min}$ & 5.0, 10.0, 25.0 & Minimum scaling factor of the radiation field intensity\\
                $\alpha$ & 2.0 & Dust power law slope\\
                $\gamma$ & 0.02 & Illuminated fraction\\
                & & \\
                \hline
                \multicolumn{3}{c}{AGN Emission}\\
                \hline
                & & \\
                $r_{ratio}$ & 60.0 & Ratio of maximum to minimum radii\\
                $\tau$ & 1.0, 6.0 & Optical depth at 9.7~$\mu$m\\
                $\beta$ & -0.5 & $\beta$ coefficient for the gas density function of the torus\tablefootmark{a}\\
                $\gamma$ & 0.0 & $\gamma$ coefficient for the gas density function of the torus\tablefootmark{a}\\
                Opening Angle & 100.0\degr & Opening angle of the torus\\
                $\psi$ & 0.001\degr, 89.990\degr & Angle between equatorial axis and line of sight\\
                $frac_{AGN}$ & 0.0, 0.1, 0.3, 0.5, 0.7 & AGN fraction\\
                & & \\
                \hline
        \end{tabular}
        \tablefoot{
                \tablefoottext{a}{Density function of the torus can be found in \citet{2006MNRAS.366..767F} as Equation 3.}
        }
\end{table*}

\section{XID+ prior width}\label{app:XID+prior}
\begin{figure}
        \centering
        \includegraphics[width=0.45\textwidth]{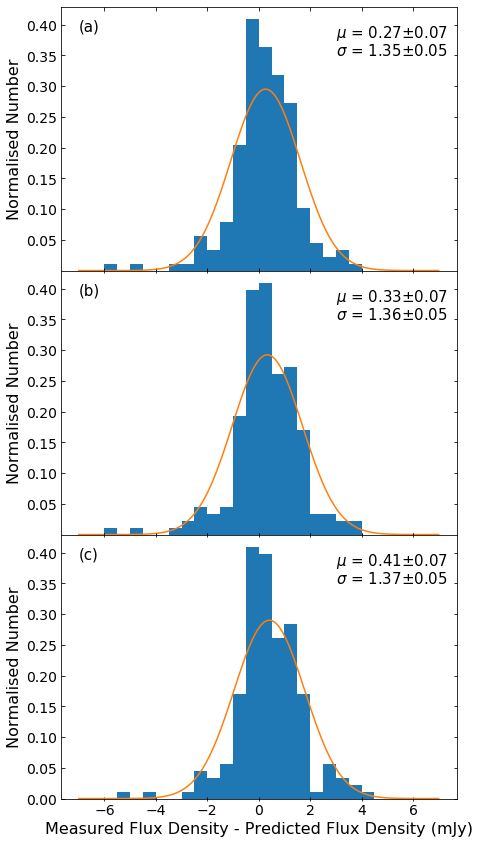}
        \caption{Distributions for the difference between the observed ALMA 870~$\mu$m flux densities and the predicted 870~$\mu$m flux densities from CIGALE using an error expansion factor of 2 (a), 3 (b), and 4 (c) in XID+. The Gaussian distribution for each expansion factor is also shown in orange, along with the means ($\mu$) and standard deviations ($\sigma$) of the distributions. All the distributions have an approximately consistent $\sigma$ so the best expansion factor was deemed to be that with $\mu$ closest to zero: 2 times the error from CIGALE.}
        \label{fig:error_width}
\end{figure}

We require the prior from CIGALE to be as informative as possible in XID+ while not being overly restrictive. To determine the preferred expansion factor of the SPIRE flux density estimate errors from CIGALE, we ran XID+ on the tiles that contain at least one of 178 objects detected at 870$\mu$m from \citet{2014ApJ...783...84S, 2016ApJ...820...83S} three times; expanding the flux density estimate error by a factor of 2, 3, and 4. The SPIRE flux densities from XID+ were then added to the data from COSMOS2015 and CIGALE was rerun to predict the Atacama Large Millimeter/submillimeter Array (ALMA) 870~$\mu$m flux densities, for each of the three XID+ results. These predictions were compared to the ALMA 870~$\mu$m observations by subtracting the CIGALE predictions from the observations and fitting a Gaussian to the resulting distribution. The histograms of the comparisons, along with the fitted Gaussian distributions, are found in Fig. \ref{fig:error_width}. As the standard deviations are all approximately consistent, we chose to expand the errors from CIGALE by a factor of two as this provides the smallest deviation of the mean from zero. 

\section{Further comparisons with \citet{2015A&A...575A..74S}}\label{app:schreiber}
As mentioned in Sect. \ref{subsec:otherwork}, the normalisation presented in this work is consistent with that of \citet{2015A&A...575A..74S} once we account for IMF and SFR tracers. We also note that the high-mass turn-over found in \citet{2015A&A...575A..74S} is not a result of stacking. We discuss this here in further detail.

\begin{figure*}
        \centering
        \includegraphics[width=0.9\textwidth]{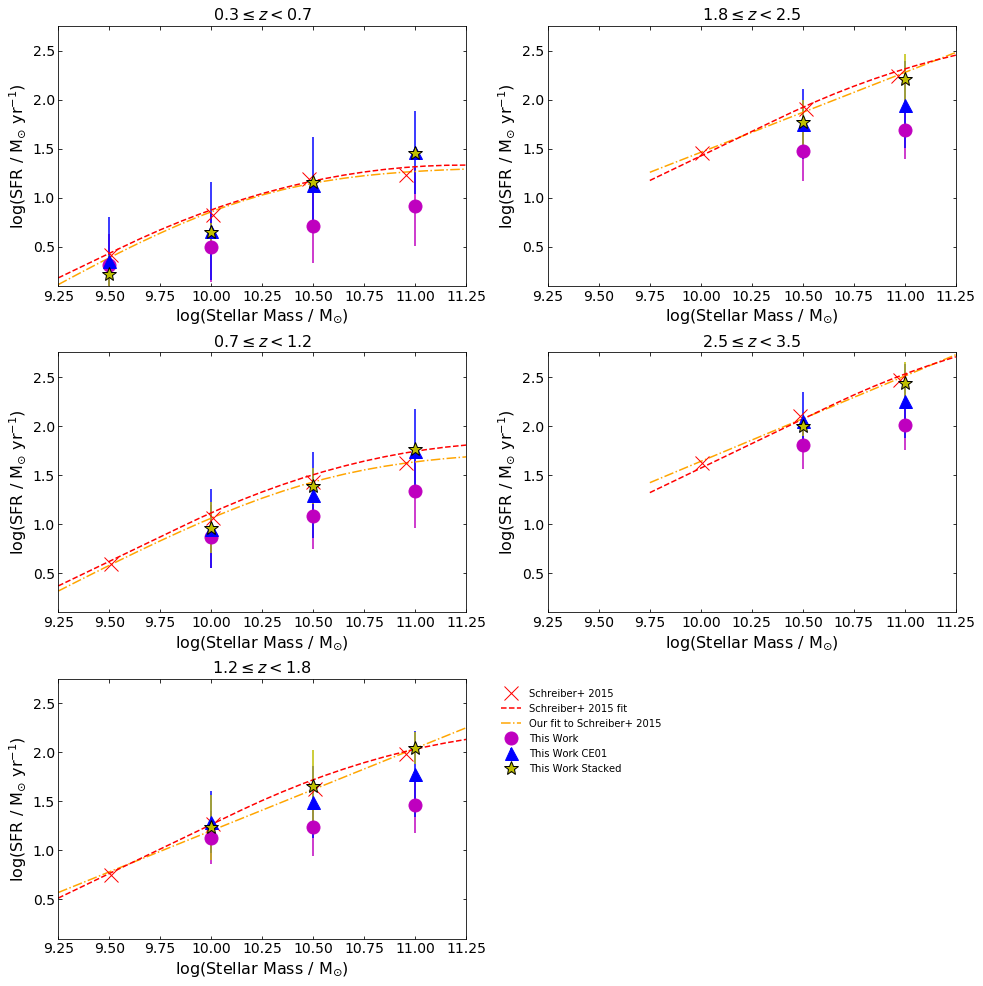}
        \caption{Comparisons of \citet[][red crosses]{2015A&A...575A..74S} SFR-M$_{\star}$ positions with redshift to different methods of deriving SFR in this work. The mean CIGALE SFR in each mass bin is in magenta, the fitting of the de-blended SPIRE data with \citet[][CE01]{2001ApJ...556..562C} templates to determine the infrared luminosity (L$_{IR}$) derived SFR is indicated with blue triangles, the IAS Stacking Library derived SFR  with yellow stars, the full \citet{2015A&A...575A..74S} MS trends with red dashed lines, and our own fits to the \citet{2015A&A...575A..74S} data with  orange dot-dashed lines. Horizontal error bars are omitted for ease of examination, but are all 0.25~dex. The \citet{2015A&A...575A..74S} data is complete to lower masses due to their use of the deeper GOODS data. As can be seen, the SFRs from SED fitting are systematically lower than those from stacking or template fitting. However, our stacked and CE01 data points are consistent with \citet{2015A&A...575A..74S} within error.}
        \label{fig:stack}
\end{figure*}

The disparity in normalisation is likely a result of the different methods of determining SFR. \citet{2015A&A...575A..74S} convert the UV and IR luminosities into UV and IR SFRs and then combine these values to generate a total SFR, while we use full SED fitting. As a result, our SFRs are lower as not all of the IR luminosity will be attributed to the young stellar population, but it also includes a component for the older stellar population. We have also fitted the \citet{2001ApJ...556..562C} templates to our own de-blended SPIRE data, calculated the IR luminosities, and converted these into SFRs. From this fitting, we recover similar SFRs to those found by \citet{2015A&A...575A..74S} (Fig. \ref{fig:stack},  blue triangles). Hence, the SFRs from SED fitting are systematically lower than those found using IR luminosity conversion.

\begin{figure}
        \centering
        \includegraphics[width=0.5\textwidth]{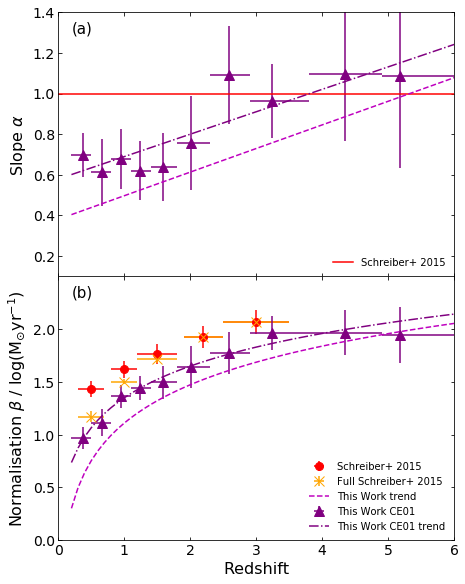}
        \caption{Comparison of the UVJ selected MS results of this work using the \citet[][CE01]{2001ApJ...556..562C} template derived SFR (dark purple triangles) with the \citet{2015A&A...575A..74S} low-mass MS (red). The $\alpha$ and $\beta$ parameters from Eq. \ref{eqn:whitaker} are in panels (a) and (b), respectively, and the trends from Fig. \ref{fig:work-comp} are in magenta. As \citet{2015A&A...575A..74S} hold their low-mass slope constant at unity, this is indicated in panel (a) as the flat red line. The orange crosses in panel (b) are the normalisations for the full mass range (including turn-over) from \citet{2015A&A...575A..74S}. The redshifts shown for this work are the mean redshift in each redshift bin while the horizontal error bars show the width of the redshift bin. The good agreement between the purple triangles and orange crosses and poor agreement between the purple triangles and red circles demonstrates how forcing a low-mass slope of unity results in normalisations that are too high at lower redshifts.}
        \label{fig:work-comp-CE01}
\end{figure}

The forward modelling was re-done using the SFRs derived from the fitting of the \citet{2001ApJ...556..562C} templates to our SPIRE data. The slope ($\alpha$) and normalisation ($\beta$) were found to evolve as
\begin{equation}
        \begin{aligned}
                \alpha(z) = (0.58 \pm 0.05) + (0.11 \pm 0.03)z,\\
                \beta(z) = (1.31 \pm 0.11) + [(0.46 \pm 0.08) \times \ln(\{0.09 \pm 0.20\} + z)].
        \end{aligned}
\end{equation}
As can be seen in Fig. \ref{fig:work-comp-CE01}, the slope using the \citet{2001ApJ...556..562C} SFR evolves with redshift at the same rate as when the CIGALE SFRs are used but the slopes are steeper, a result of high-mass galaxies having more older stars \citep[e.g.][]{2005MNRAS.362...41G}, which will lower the CIGALE SFR with respect to the \citet{2001ApJ...556..562C} SFR for these high-mass objects. As expected, the normalisations are closer to those of the \citet{2015A&A...575A..74S} low-mass slope and become consistent, within error, at high redshift.

The remaining normalisation difference at low redshift is a result of the enforced slope of unity in \citet{2015A&A...575A..74S}. We fit Eq. \ref{eqn:whitaker} to the data in \citet{2015A&A...575A..74S} Fig. 10 and find slopes below unity. As a result of this fitting, the normalisations from our fits to the \citet{2015A&A...575A..74S} data agree with the normalisations we find when fitting to our IR luminosity derived SFRs. Figure \ref{fig:work-comp-CE01}b also includes the normalisations for the full mass range from \citet{2015A&A...575A..74S}, including their high-mass turn-over, as orange crosses. These normalisations are in good agreement with the normalisations found in this work, demonstrating how enforcing a low-mass slope of unity results in normalisations that are too high at lower redshifts.

Our own fitting to the \citet{2015A&A...575A..74S} MS also provides a slightly different interpretation of their data. In addition to  fitting  Eq. \ref{eqn:whitaker} to their data, we also fit Eq. \ref{eqn:lee}. The result of this refitting is that we only find evidence of a high-mass turn-over in the two lowest  redshift bins used by  \citet{2015A&A...575A..74S};  all the other redshift bins are consistent with a simple power law (Fig. \ref{fig:stack} red dashed lines). Hence, their high-mass turn-over above $z = 1.2$ may be a result of forcing a low-mass slope of unity.

To explore whether our lack of a high-mass turn-over at low redshift is due to stacking, we stacked the SPIRE \citep[from HerMES,][]{2012MNRAS.424.1614O} and PACS \citep[from PACS Evolutionary Probe,][]{2011A&A...532A..90L} images of the COSMOS field on our source positions, using the same UVJ selection, mass bins, and redshift bins as \citet{2015A&A...575A..74S} to generate a fair comparison. We used two independent stacking codes: the IAS Stacking Library \citep{bavouzet:tel-00363975, 2010A&A...512A..78B} for luminosity stacking (Fig. \ref{fig:stack} yellow stars), and Simstack \citep{2013ApJ...779...32V} for flux density stacking. The stacked \textit{Herschel} flux densities were then fitted with the \citet{2001ApJ...556..562C} IR templates to calculate the IR luminosity and SFR. We find no evidence of a high-mass turn-over using either of the stacking tools.

\end{appendix}

\end{document}